\documentclass[aps,pre,onecolumn,floatfix]{revtex4}

\usepackage{graphicx, bm, bbm,amsmath, amssymb, epsfig}

\newcommand{\nn}{\noindent}

\newcommand{\bq}{\begin{align}}
\newcommand{\eq}{\end{align}}

\begin{document}

\title{Packing of elastic rings with friction}

\author{
Silas Alben}

\affiliation{Department of Mathematics, University of Michigan,
Ann Arbor, MI 48109, USA}
\email{alben@umich.edu}

\date{\today}

\begin{abstract}
We study the deformations of elastic filaments confined within slowly-shrinking circular boundaries, under contact forces with friction. We perform computations with a spring-lattice model that deforms like a thin inextensible filament of uniform bending stiffness. Early in the deformation, two lobes of the filament make contact. If the friction coefficient is small enough, one lobe slides inside the other; otherwise, the lobes move together or one lobe bifurcates the other. There follows a sequence of deformations that is a mixture of spiraling and bifurcations, primarily the former with small friction and the latter with large friction. With zero friction, a simple model predicts that the maximum curvature and the total elastic energy scale as the wall radius to the -3/2 and -2 powers respectively. With nonzero friction, the elastic energy follows a similar scaling but with a prefactor up to 8 times larger, due to delayering and bending with a range of small curvatures. For friction coefficients as large as 1, the deformations are qualitatively similar with and without friction at the outer wall. Above 1, the wall friction case becomes dominated by buckling near the wall.
\end{abstract}


\maketitle

\section{Introduction}

There have been many studies in recent years of the packing and crumpling of thin elastic sheets and rods under confinement, using 
theory \cite{cerda2005confined,boue2007folding,conti2008confining,deboeuf2009energy,adda2010statistical,bayart2011measuring,oshri2015wrinkles,elettro2017elastocapillary,andrejevic2021model}, computations \cite{vliegenthart2006forced}, and experiments \cite{matan2002crumpling,cambou2011three}. Many studies have examined the basic physics and mechanics of the packing process \cite{donato2003scaling,boue2006spiral,donato2007condensation,gomes2008plastic,stoop2008morphological,lin2008crumpling,gomes2010crumpled,stoop2011packing,najafi2012ordered,pineirua2013spooling,vetter2014morphogenesis,vetter2015packing,napoli2015snap,sobral2015unpacking,sobral2015tight,box2020dynamic,grossman2021packing}, and the geometry of the buckled and creased shapes \cite{lobkovsky1995scaling,lobkovsky1997properties,alben2007self,muller2008conical,katifori2010foldable,davidovitch2011prototypical,alben2011edge,deboeuf2013comparative,alben2015bending,paulsen2016curvature,hoffmann2019machine,alben2019semi,tobasco2020exact}. Other studies have considered biological applications \cite{rim2014mechanical} such as packing of chromosomal DNA inside cell nuclei \cite{yoo2014physics,lankavs2020simple}, and packing of viral RNA and DNA inside a protein capsid container \cite{tzlil2003forces,klug2005three}. Both elastic and plastic deformations have been considered. 
\noindent A particular problem of interest has been the 2D deformations of thin elastic rings under forcing by external contacts or pressure \cite{lo1962buckling,chan1966stability,el1972buckling,kyriakides1984collapse,hazel2017buckling,katifori2009collapse,yang2019buckling}; this is also the focus of the present paper.

\begin{figure}[!h]
\hspace*{-0.2in}
\includegraphics[width=3in]{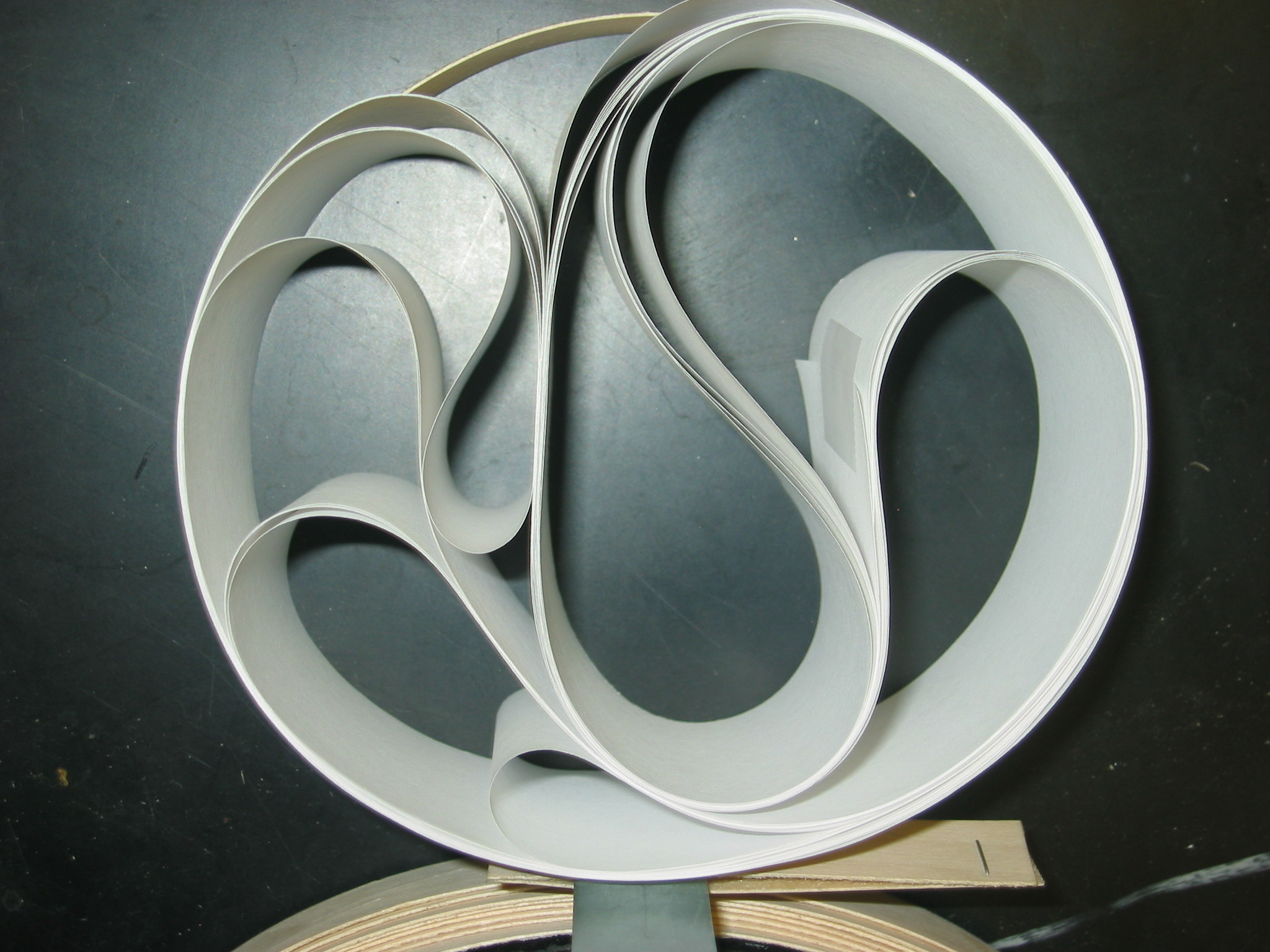}\includegraphics[width=3in]{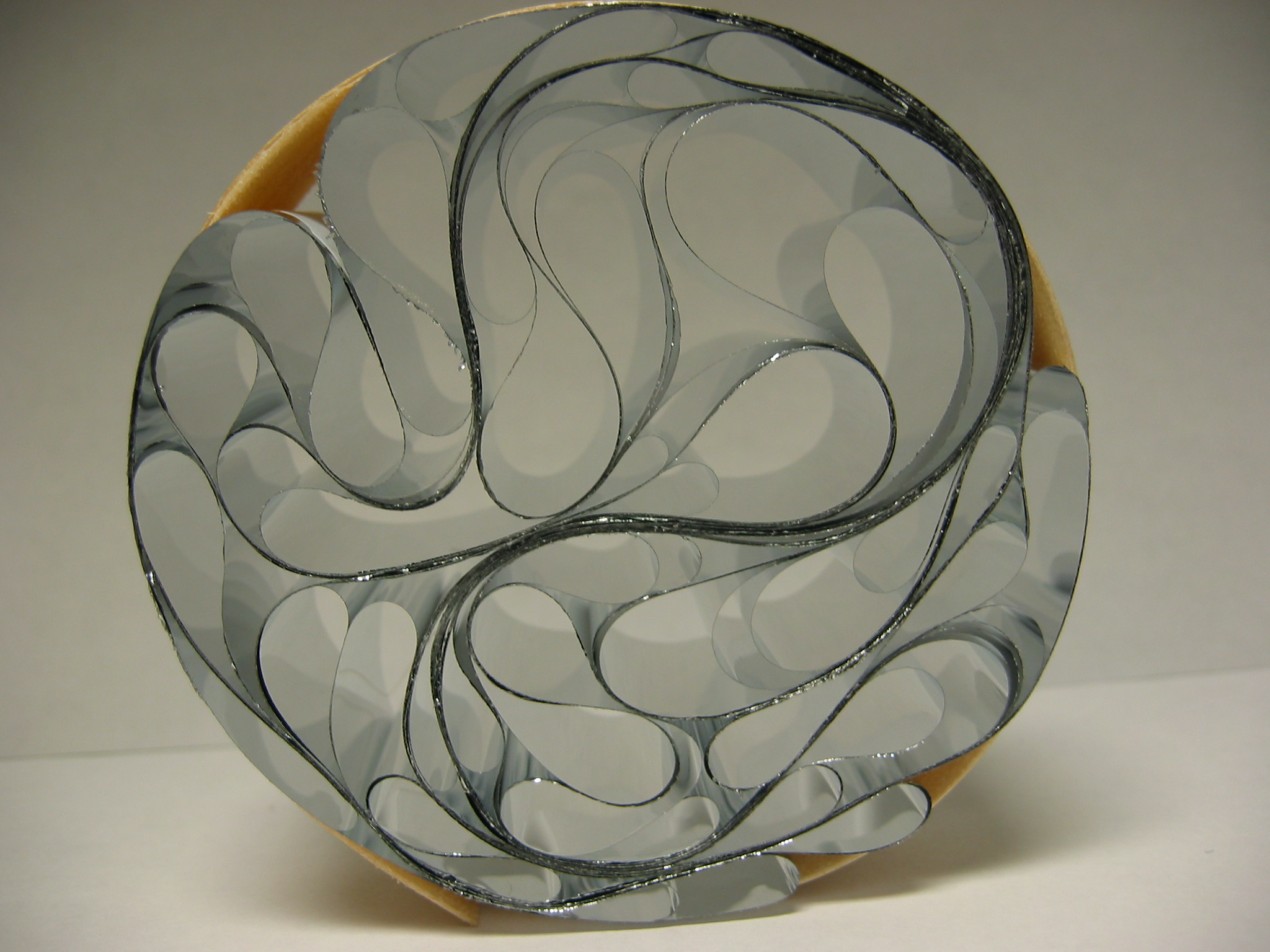}
\caption{Examples of paper (left) and mylar rings (right) compressed by a gradually shrinking outer boundary that is a thin strip of wood. The outer boundary shrinks as one end of the wood strip passes through a loop, in the same way that a clothing belt is tightened. Right panel reproduced from \cite{spears2008cascade}, with the permission of AIP Publishing.}
\label{fig:ChaosExpt}
\end{figure}

This paper studies a basic version of the problem in which an initially circular elastic ring is compressed by a shrinking circular container. In an experiment \cite{spears2008cascade}, a system of contact forces between different points on the ring, and between the ring and the confining boundary led to a complex and heterogeneous configuration with a wide range of bending length scales and the spontaneous layering of the ring, with different numbers of layers formed at different locations. An inverse relationship between the number of layers and the curvature at a given location was identified. In this experiment (see Fig. \ref{fig:ChaosExpt}), the confining boundary was a strip of wood, thin enough to be flexible but much more rigid than the confined ring, which is a thin strip of paper (panel A) or mylar (panel B). The radius of the circular wood boundary was gradually decreased in the same way that a belt is tightened. A similar process was studied in a slightly different configuration by passing a long conical sheet through a hole \cite{boue2006spiral}. This study focused particularly on the case of low friction, in which most of the ring eventually forms a spiral, except for a single sharp bend located outside the spiral.

The goal of the present work is to study the effect of friction on the packing process. We use numerical simulations of a model based on Coulomb friction. Although the model is somewhat idealized, this approach allows for precise control of the packing process and high spatial resolution of the resulting structures. 

\section{Model}

We will study a thin ring of elastic material---a closed filament---with thickness $h \ll L$, its length. We assume that the ring only deforms in the 2D plane, but has width $W$ in the out-of-plane direction. The ratio of the stretching modulus ($EhW$) and bending modulus ($Eh^3W/12$) is $O\left(h^{-2}\right)$, so in the limit of small thickness, the filament deforms by bending only, without stretching, in response to generic external forces.

\begin{figure}[!h]
\hspace*{-0.2in}
\includegraphics[width=6.2in]{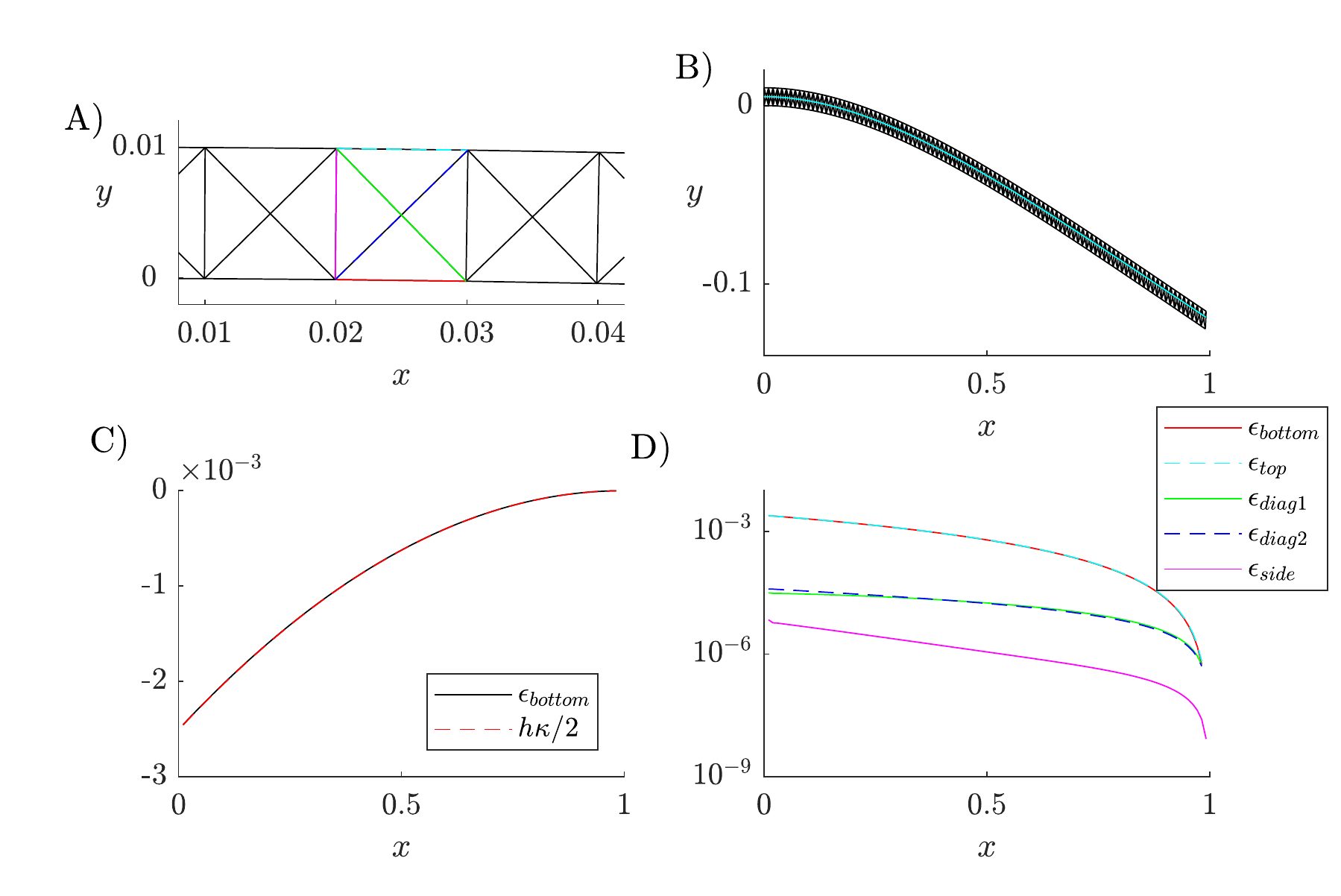}
\caption{X-lattice spring model of a bending beam. A) Close-up of three square units near the clamped end of the beam ($x = 0$), when the deformation is very small. Certain springs are colored, corresponding to quantities plotted in panel D. B) Comparison between X-lattice solution to cantilever problem (black lines) and nonlinear continuum solution (thin green line close to the centerline of the X-lattice). C) Comparison of the strain in the bottom springs of each X-lattice element (black line) and the continuum formula $h\kappa/2$ (red dashed line). D) Distributions of strains in each of the five classes of springs in the lattice elements (bottom, top, both diagonals, and side) versus the $x$-coordinates of the midpoints of the springs.}
\label{fig:CantileverFig}
\end{figure}

We compute deformations of the thin elastic ring using a discrete model that behaves like an isotropic continuum material on scales much larger than that of the discretization. The thin ring is represented as a thin lattice of springs, consisting of a chain of square units in which Hookean springs connect each of the vertices in a given square with the other three vertices (see Fig. \ref{fig:CantileverFig}A). Spring-lattice models of continuum elastic rods and plates are well established in the literature \cite{ostoja2002lattice, hrennikoff1941solution,noor1988continuum}. Here the springs have rest lengths $h$ (along the sides of the squares) or $\sqrt{2}h$ (along the diagonals), equal to their lengths in the undeformed squares. We use only stretching springs, rather than a combination of stretching and bending springs as in other recent works \cite{lobkovsky1995scaling,alben2007self,katifori2009collapse}. The condition number of the system of equations in our energy minimization routine scales as $h^{-2}$ with stretching springs only, versus $h^{-4}$ with bending springs. Using only stretching springs allows us to go to smaller $h$ before ill-conditioning prevents convergence, and we can then simulate smaller-scale deformations.

We consider as a test problem an inextensible cantilever beam of unit length, clamped at one end ($s = 0$; $s$ is the arc length coordinate along the beam centerline) and free at the other end ($s = 1$). The force balance equation is
\begin{align}
    -\partial_s(B \partial_s \kappa \hat{\mathbf{n}}) + \partial_s(T \hat{\mathbf{s}}) + \mathbf{f}_{ext} = 0. \label{cantilever}
\end{align}
\nn where $B$ is the uniform bending modulus, $\kappa(s)$ is the curvature, and $T(s)$ is the internal tension that prevents extension. We take the applied external force to be
$\mathbf{f}_{ext} = -\hat{\mathbf{e}}_y$, a unit downward force per unit length.
The $\hat{\mathbf{s}}$-component of (\ref{cantilever}) can be integrated to solve for the tension:
\begin{align}
    T(s) = -\frac{B}{2}\kappa^2 + y(s) - y(1)
\end{align}
\nn which is then inserted into the $\hat{\mathbf{n}}$-component of (\ref{cantilever}):
\begin{align}
    -B\partial_{ss}\kappa + T\kappa -\cos{\theta} = 0. \label{fn}
\end{align}
\nn where $\theta(s)$ is the tangent angle of the beam. We solve (\ref{fn}) by the method of successive approximations, starting with the flat state $y(s) = \theta(s) = \kappa(s) \equiv 0$ as an initial guess. At each iteration, (\ref{fn}) is integrated starting from the free end using a second-order Runge-Kutta method with boundary conditions $\partial_s\kappa(1) = \kappa(1) = 0$.
With a uniform $s$-grid with spacing 0.0005, convergence to $O(10^{-16})$ occurs in about 10 iterations. The solution with $B = 1$ is shown by the green line in Fig. \ref{fig:CantileverFig}B. 

We now solve the same problem with the X-lattice model, using 100 square units and $h = 0.01$. We write the elastic energy of the cantilever:
\begin{align}
    U_{elastic} = \frac{k_1}{2} \sum_{i,j} \left(\|\mathbf{X}_i - \mathbf{X}_j\| - d_{ij} \right)^2 \label{Uelastic}
\end{align}
\nn where $d_{ij}$ is either $h$ or $\sqrt{2}h$, the rest length of the spring connecting points $\mathbf{X}_i$ and $\mathbf{X}_j$. We solve the equilibrium equations 
$\nabla U_{elastic} + F_{ext} = 0$, where $F_{ext}$ is a discrete approximation to a unit downward force per unit length, applied half to the top row of points in Fig. \ref{fig:CantileverFig}A and half to the bottom row. In appendix \ref{k1} we show that $B = k_1 h^3/2$ gives the bending modulus $B$ that corresponds to spring stiffness $k_1$, in the limit of small $h$, when the X-lattice behaves approximately as an Euler-Bernoulli beam. Using Newton's method to solve $\nabla U_{elastic} + F_{ext} = 0$ with $B = 1$ (i.e. $k_1 = 2h^{-3}$), 
we obtain the solution shown by the black lines in Fig. \ref{fig:CantileverFig}B, compared with the green line for the continuum model. In panel C, we find that the strain in the bottom springs agrees with the continuum formula $h\kappa/2$, and in panel D, we show the strain in the five different types of springs along the length of the cantilever. The strain in the top and bottom springs is larger than that in the other springs by about a factor of $1/h$ (100 here) over most of the lattice, so the deformation is primarily bending with little shear.

\begin{figure}[!h]
\hspace*{-0.2in}
\includegraphics[width=6.2in]{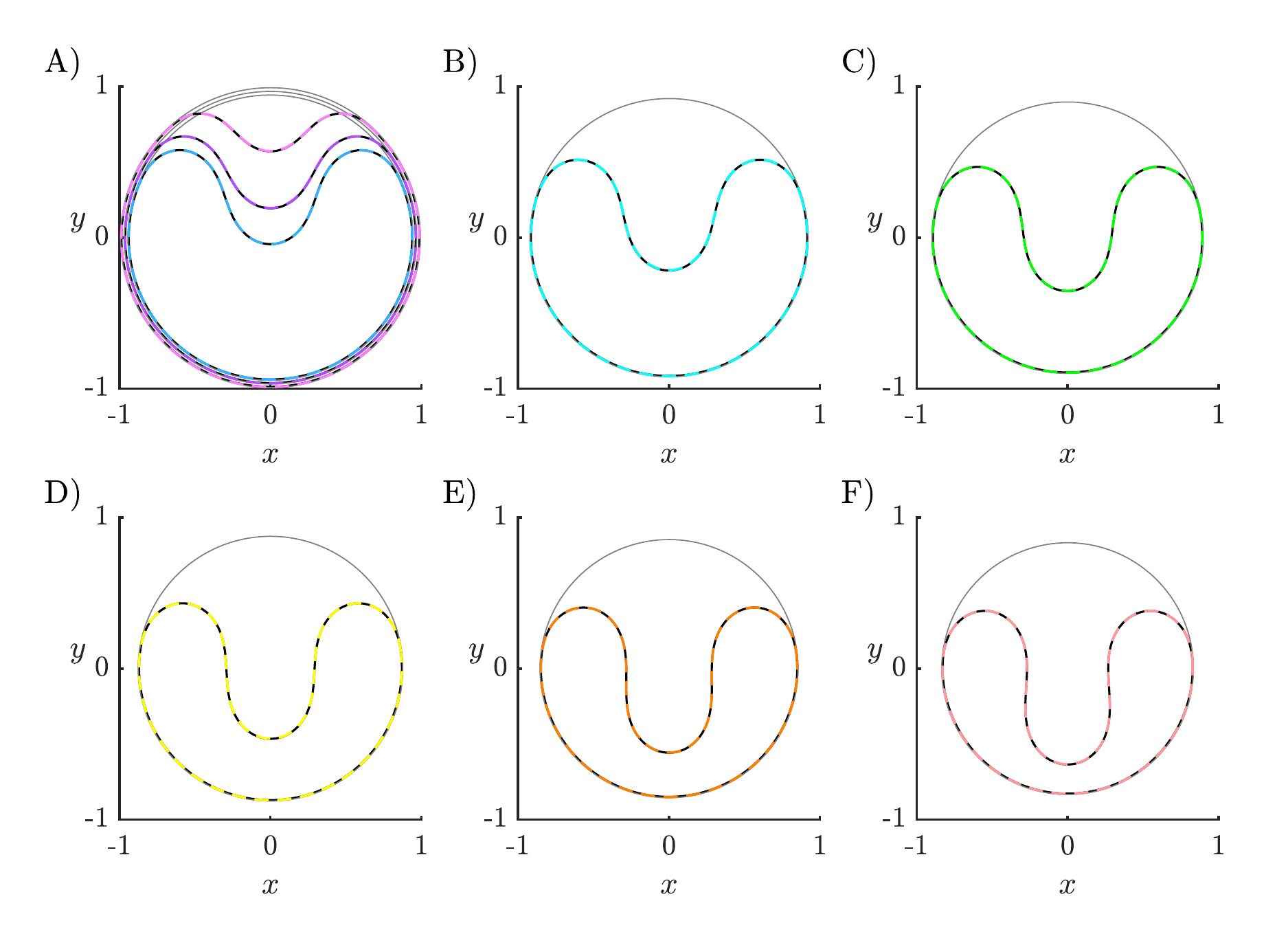}
\caption{A comparison of X-lattice solutions with $n$ = 1000 (black lines) and inextensible elastica solutions (colored dashed lines) with the same maximum radii. The confining rings of the X-lattice beams are gray circles with radii $R_w$ given by A) 0.993, 0.968, 0.944, B) 0.921, C) 0.898, D) 0.876, E) 0.855, and F) 0.0.833.}
\label{fig:CompareAnalyticalRing}
\end{figure}

We now make a second comparison between the X-lattice and the continuum model, for the beginning stages of the problem considered in this paper, an elastic ring confined within a shrinking circular boundary. We connect the ends of the top row of points in Fig. \ref{cantilever}A by springs and likewise for the bottom row. In the absence of external forces, the two rows are now inner and outer circular rings in the simplest equilibrium state. Because the lattice is now curved, it has nonzero elastic energy; the inner ring is compressed and the outer ring is extended. 
With $n$ points in each ring, we take $h$ = $2\pi/n$, giving a circle of length 2$\pi$ and radius 1 in the limit of small $h$ (or large $n$) in the unconfined equilibrium state. 

When the radius of the confining boundary is decreased to slightly less than 1, another equilibrium state---a ``puckered ring''---occurs for sufficiently thin rings. It is exemplified by the outermost filament (light purple) in Fig. \ref{fig:CompareAnalyticalRing}A. The shape can be divided into two regions---a circular region that is in contact with the outer boundary, and a puckered region that is buckled inward.
The problem has been studied by a number of works \cite{lo1962buckling,chan1966stability,el1972buckling,cerda2005confined}. The curvature in the puckered region scales inversely with its length $l$, so the bending energy of the puckered region, $\sim \int \kappa^2 ds \sim l^{-2} l = l^{-1}$, diverges as the pucker becomes {\it smaller}. If the ring is inextensible (the limiting case of zero thickness), the puckered state is the only equilibrium when the boundary radius is slightly below 1.  

In \cite{chan1966stability}, analytical solutions for the puckered region deflection were given in terms of elliptic integrals, using Euler's elastica equation. They showed that for an {\it extensible} ring with a small but nonzero thickness, as the outer boundary radius decreases from the equilibrium radius of the unconfined ring (unity), there is a range of boundary radii near unity for which the uniformly compressed circular state is the only equilibrium. Then, at a critical boundary radius whose distance from unity depends on the ring thickness, a branch point appears with two additional branches of equilibria, one stable (e.g. the puckered rings in Fig. \ref{fig:CompareAnalyticalRing}A) with lower energy than the compressed circle, and one unstable and with higher energy than the compressed circle. A finite perturbation is needed to move the ring over the energy barrier from the compressed circular state to the puckered state. The size of the perturbation decreases with decreasing ring thickness. 

In Fig. \ref{fig:CompareAnalyticalRing},  X-lattice solutions of the puckered ring for $n = 1000$ are shown by black lines. Superposed on them are colored dashed lines, solutions to Euler's elastica equation in the puckered region. To compute the puckered X-lattice solutions, we use Newton's method to minimize an energy that is the sum of the elastic energy (\ref{Uelastic}) and a wall energy that prevents the ring from penetrating the outer boundary:
\begin{align}
    U_{wall} &= K_w \sum_{j = 1}^{2n} F\left(\frac{\|\mathbf{X}_j\|-R_w}{\delta_w}\right) \\
    F(r) &= \begin{cases}
       e^{r}, & r < 0 \\
     1 + r + r^2/2, & r \geq 0.
    \end{cases}   \label{Uwall}   
\end{align}
\nn The exponential fall-off with negative $r$ ensures that the energy is insignificant when points are much farther than $\delta_w$ from the wall, here taken as 
$1.5 \times 10^{-5}$. $F$ changes smoothly to a polynomial behavior when $r \geq 0$, to avoid computing with extremely large values of the exponential when $r \gg \delta_w$. We take $K_w = 10^{-5}$ so that the wall force becomes comparable to the bending force when $r$ is a small multiple of $\delta_w$.
We guide the ring from the compressed circular state to the puckered state by temporarily introducing a spring between a point on the ring and a fixed point closer to the center of the ring. This causes an indentation at this point when the energy is minimized. When this temporary spring is removed and the energy is minimized again, the indented shape becomes the smooth puckered shape.

The elastica solutions
are computed with a second-order Runge-Kutta method with spatial grid size $2\pi \times 10^{-4}$. 
Our method for generating the full range of elastica solutions with a single puckered region (including those in Fig. \ref{fig:CompareAnalyticalRing}) is described in appendix \ref{pucker}.

As $R_w$ decreases below 0.82, the puckered region contacts the portion of the elastic ring along the bottom of the outer boundary. It is possible to track the self-contacts and solve the elastica equation for the elastic ring shape as in \cite{boue2006spiral}, but it becomes increasingly complicated as the number of self-contacts increases. Therefore, we proceed with the X-lattice model only, first studying the case of zero friction, which was found to result in a spiral configuration \cite{boue2006spiral}.

\section{Packing without friction}

In order to prevent self-penetration of the ring, we introduce another repulsive term in the energy, analogous to the wall repulsion (\ref{Uwall}):
\begin{align}
    U_{self} &= K_w \sum_{i = 1}^{2n} F\left(\frac{n_{i}}{\delta_w}\right).\label{Uself}   
\end{align}
\nn Here $n_{i}$ is the 
normal component of the displacement from point $i$ to any point on the nearest non-neighboring segment of the elastic ring. $F$ and $\delta_w$ are defined as before. By ``segment'' we mean a line segment between two adjacent points on the inner or outer elastic rings. We determine the ``nearest non-neighboring segment'' by excluding neighboring adjacent segments (which may have very small normal distance, being almost tangentially displaced from point $i$), so only a segment that is nearly in normal contact with point $i$ is selected. In practice this is achieved by first determining the nearest point to point $i$ from among all points except $i$ and its eight nearest neighbors in either direction along the ring. If the nearest point thus found is closer than $5h$ from point $i$, it is close to a state of normal contact with $i$, because the only the eight nearest neighbors are closer than $8h$ (assuming the ring is nearly straight between near neighbors, i.e. the curvature is resolved well by the discretization) and these are excluded. The ``nearest segment'' is then taken to be that connecting the nearest point just defined, and the closest of its nearest neighbors to point $i$.

\begin{figure}[!h]
\hspace*{-0.2in}
\includegraphics[width=6.2in]{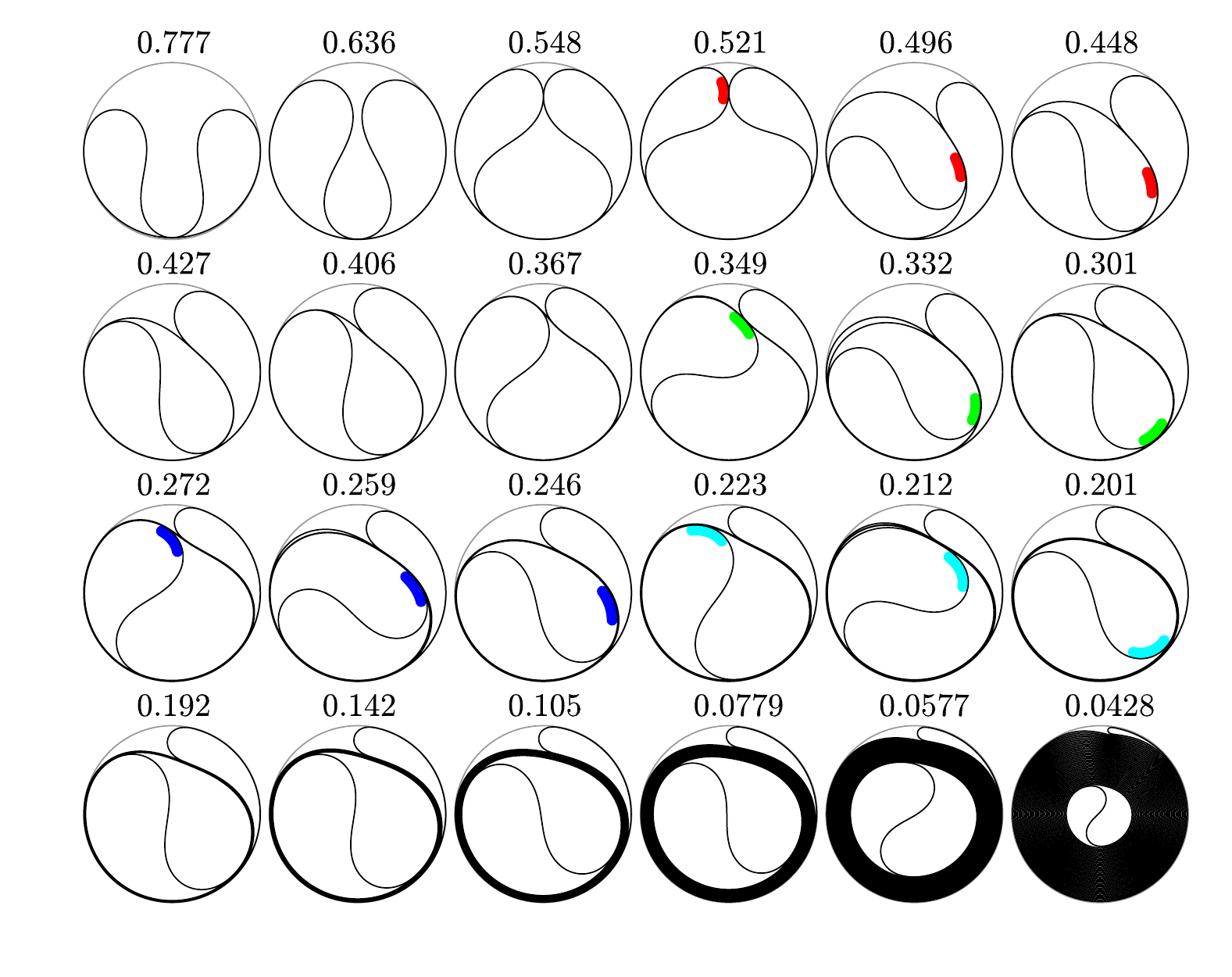}
\caption{X-lattice solutions with zero friction and $n$ = 10000 at various outer boundary radii $R_w$ (labeled above each ring). For $0.201 \leq R_w \leq 0.521$, colors are used to label fixed sections of material in three successive configurations.}
\label{fig:ZeroFrictionFig}
\end{figure}

We now increase $n$ to 10000, to be able to resolve the larger curvatures that arise with a greater degree of packing. In Fig. \ref{fig:ZeroFrictionFig} we show the X-lattice solutions as $R_w$ decreases from 0.777 (upper left), where the puckered region contacts the ring along the bottom boundary, to 0.0428 (lower right), the most confined state. Here most of the elastic ring has assumed the shape of a spiral with 35 turns, except for a small sharp bend outside the spiral, adjacent to the wall. Somewhat below this $R_w$, discretization effects become important: a sharp angular feature abruptly appears near the sharp bend in the energy-minimizing state, which therefore no longer approximates a smooth curve. We have scaled the outer boundary to the same size in each case, to make the ring configuration more visible. We again use Newton's method to minimize the energy here, given by
$U_{elastic} + U_{wall} + U_{self}$. We use the continuation method to compute solutions across a range of $R_w$: we decrease $R_w$ in a sequence of small steps (multiplying $R_w$ by 0.9998 at each step), and use the energy minimizer at the previous $R_w$ as the initial guess for the minimizer at the next $R_w$.

The initial steps in the formation of the spiral, which agree with the elastica solutions in \cite{boue2006spiral}, are shown in the top row. At $R_w$ = 0.548, two symmetric lobes make contact at the top of the ring. The force between the two increases until an asymmetric state emerges---one lobe slides inside the other at $R_w$ = 0.496. A section of material is labeled in red as $R_w$ decreases from 0.521 to 0.448 to illustrate the lobe motion. The red section flattens, indicating that the inner lobe slides and also rolls along the contacting surface, so that the curvature maximum moves forward with respect to the fixed material shown in red. Almost the same process occurs again in the second row, as the inner S curve rotates by about a half-turn, adding a layer onto the spiral. An inner lobe again slides past the outer lobe as $R_w$ decreases from 0.367 to 0.301, shown by the material marked in green. The lobes are more asymmetric now, and there is a layer of material between them. Two later cases of the same process are shown in the third row, with two regions of material marked by dark blue and light blue respectively. The bottom row shows later stages of the deformation, when the spiral has 5, 7, 10, 14, 20, and 35 turns, respectively. 

\begin{figure}[!h]
\hspace*{-0.2in}
\includegraphics[width=6.2in]{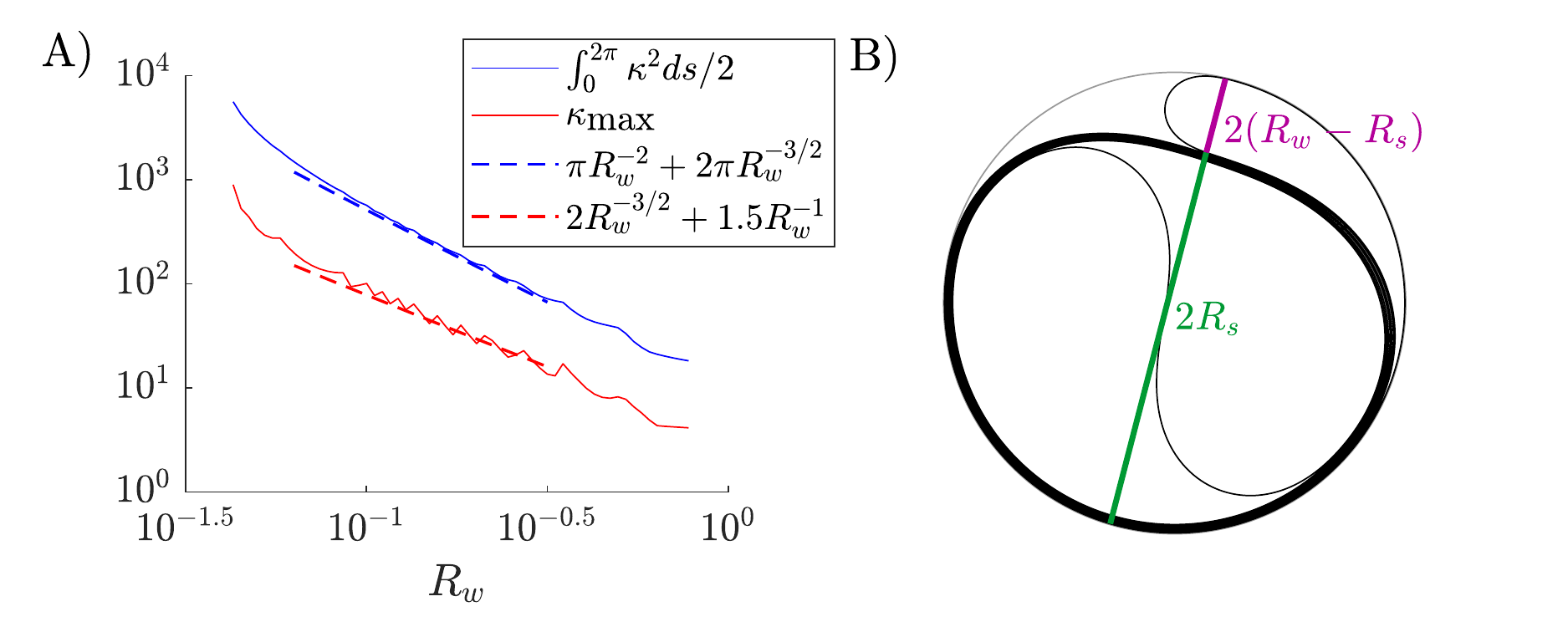}
\caption{A, Total elastic energy (solid blue line) and maximum curvature (solid red line) together with asymptotic approximations (dashed blue and red lines). B) Definition of the spiral diameter $2 R_s$ and the remainder of the boundary diameter, $2(R_w-R_s)$.}
\label{fig:ZeroFrictionAnalysis}
\end{figure}

In Fig. \ref{fig:ZeroFrictionAnalysis}A we plot the total elastic energy (solid blue line) and the maximum ring curvature (solid red line) as $R_w$ decreases through the range shown in Fig. \ref{fig:ZeroFrictionFig}. The curves approximately fit two asymptotic scaling laws shown by dashed lines over an intermediate range of $R_w$, where $10^{-1.2} \lesssim R_w \lesssim 10^{-0.5}$ small enough that a spiral with multiple turns has formed, but large enough that the thickness of the ring is negligible (e.g. not for the last two shapes in Fig. \ref{fig:ZeroFrictionFig}, where the thickness increases the confinement and thus the elastic energy of the inner spiral turns). In the intermediate range of $R_w$, we approximate the shape as two regions of constant curvature. One is a semicircle of sharp curvature, which approximates the small sharp curve (in black) to the left of the small purple line in Fig. \ref{fig:ZeroFrictionAnalysis}B. The other region is the rest of the elastic filament (also black), which approximately follows the circle with diameter $2R_s$ shown by the green line in Fig. \ref{fig:ZeroFrictionAnalysis}B. The main deviation from the circle is the inner S-shaped curve, which has curvature magnitudes that are both greater and less than those of the outer spiral turns ($\approx 1/R_s$). The S-curve is just a small portion of the filament, and its curvature magnitudes are comparable to $1/R_s$, so we approximate them by that constant value. The total elastic energy in the intermediate-$R_w$ regime is thus the sum of that of the small semicircle and the rest of the filament:
\begin{align}
    U_{approx} &= \frac{B}{2}\int_0^{2\pi} \kappa^2 ds = \frac{B}{2}\left( (\kappa^2 L) |_{semicircle} + (\kappa^2 L) |_{remainder}\right)\\
    &= \frac{B}{2} \left(\frac{1}{(R_w-R_s)^2} \pi (R_w-R_s) + \frac{1}{R_s^2} (2\pi -\pi(R_w-R_s)) \right). \label{Uapprox}
\end{align}
\nn where $L$ is the length of each constant curvature portion. We set the bending modulus $B$ to 1 and assume $2\pi -\pi(R_w-R_s) \approx 2\pi$. We minimize the energy by setting $\partial_{R_s} U_{approx}$ to 0 and obtain 
\begin{align}
    R_w - R_s \approx \frac{R_s^{3/2}}{2} \longrightarrow R_s \approx R_w - \frac{R_w^{3/2}}{2} + O(R_w^2)
    \label{RwRs}
\end{align}
\nn so in the limit of small $R_w$, $R_s \sim R_w$ as expected (the spiral fills most of the domain) and $\pi(R_w-R_s) = O\left(R_w^{3/2}\right) \ll 2\pi$, consistent with the assumption.
Using (\ref{RwRs}) we write the first two terms in the asymptotic expansions of the elastic energy and the maximum curvature (that of the small semicircle):
\begin{align}
    U_{approx} &\approx \pi (R_s^{-2} + R_s^{-3/2}) = \pi (R_w^{-2} + 2 R_w^{-3/2}) +O\left(R_w^{-1}\right) \label{Uapprox1} \\
    \kappa_{max} &= \frac{1}{R_w-R_s} \approx 2R_s^{-3/2} = 2 R_w^{-3/2} + \frac{3}{2} R_w^{-1}
    + O\left(R_w^{-1/2}\right)
\end{align}
\nn The two-term approximations are shown by the blue and red dashed lines in Fig. \ref{fig:ZeroFrictionAnalysis}A, respectively. The agreement is good, but worsens as $R_w$ drops below 10$^{-1.2}$, where the filament thickness becomes important, forcing the inner spiral turns further inward, increasing their curvature and elastic energy. \cite{boue2006spiral} showed that the energy has an approximately logarithmic behavior as the spiral fills the domain. Our approximate energy minimization can be extended to this case but we do not pursue it here.



\section{Packing with friction}

We now incorporate friction into the model. This is done by generalizing (\ref{Uself}) to
\begin{align}
    U_{self} &= K_w \sum_{i = 1}^{2n} F\left(\frac{n_{i} + \mu \sqrt{s_i^2 + \delta_s^2}}{\delta_w}\right).\label{Uself1}   
\end{align}
\nn Here $n_i$ is again the normal component of the displacement of point $i$ from its nearest non-neighboring segment, or more concisely, ``contact segment.'' 
$\mu$ is the coefficient of static friction, and (\ref{Uself1}) reduces to (\ref{Uself}) when $\mu = 0$. $s_i$ is the tangential component of the displacement of point $i$ from the ``contact point.'' The contact point is the point on the contact segment that is nearest to point $i$ at the previous energy minimizer (the converged solution at the previous $R_w$ value). Taking the derivatives of (\ref{Uself1}) with respect to $n_i$ and $s_i$, we have a relation between the tangential force $f_{s,i}$ and the normal force $f_{n,i}$ of the contact:
\begin{align}
f_{s,i} = -\partial_{s_i} U_{self} = -\mu \partial_{n_i} U_{self} \frac{s_i}{\sqrt{s_i^2 + \delta_s^2}} = \mu f_{n,i}
\frac{s_i}{\sqrt{s_i^2 + \delta_s^2}}. \label{fsi}
\end{align}
\nn Therefore, in magnitude, the tangential force due to the contact is less than or equal to $\mu$ times the normal force, as required by the Coulomb law for static friction \cite{bhushan2000modern}. Here $\delta_s$ is a small constant, $10^{-8}$. By decreasing $R_w$ in a sequence of small steps (by factor of 0.9998), we are simulating the filament deformation as a quasistatic process. We wish to find a sequence of static equilibria, each one close to the previous one. In each equilibrium state, the net force is zero at each point. At a contact point, this means the sum of the elastic and the contact forces is zero. The normal component of the elastic force is equal to the negative of $\partial_{n_i} U_{self}$, the normal contact force, and
the tangential component of the elastic force is equal to the negative of $\partial_{s_i} U_{self}$, the tangential contact (i.e. frictional) force. The contact will slide when the tangential component of the elastic force is too large to be balanced by friction, and as it slides, the tangential elastic force decreases until it can be balanced again by friction. If nonnegligible sliding has occurred, $s_i \gg \delta_s$, and we
have $|f_{s,i}| \approx \mu |f_{n,i}|$ by (\ref{fsi}). That is, the frictional force is approximately as large as it can be given the normal force, and acts in the direction opposite to the sliding motion. At a static equilibrium we may have 
$|f_{s,i}| \in [0, \mu |f_{n,i}|]$, but by requiring $|f_{s,i}|$ near the upper end of the range when nonneglible sliding has occurred, we approximately minimize the amount of sliding. This is desirable because we are searching for a new static equilibrium that is close to the previous one. If instead $s_i \sim \delta_s$, the contact remains approximately static from the previous equilibrium to the new one, and then $|f_{s,i}|$ may take any value in $[0, \mu |f_{n,i}|]$ by (\ref{fsi}).

\begin{figure}[!h]
\hspace*{-0.2in}
\includegraphics[width=6.2in]{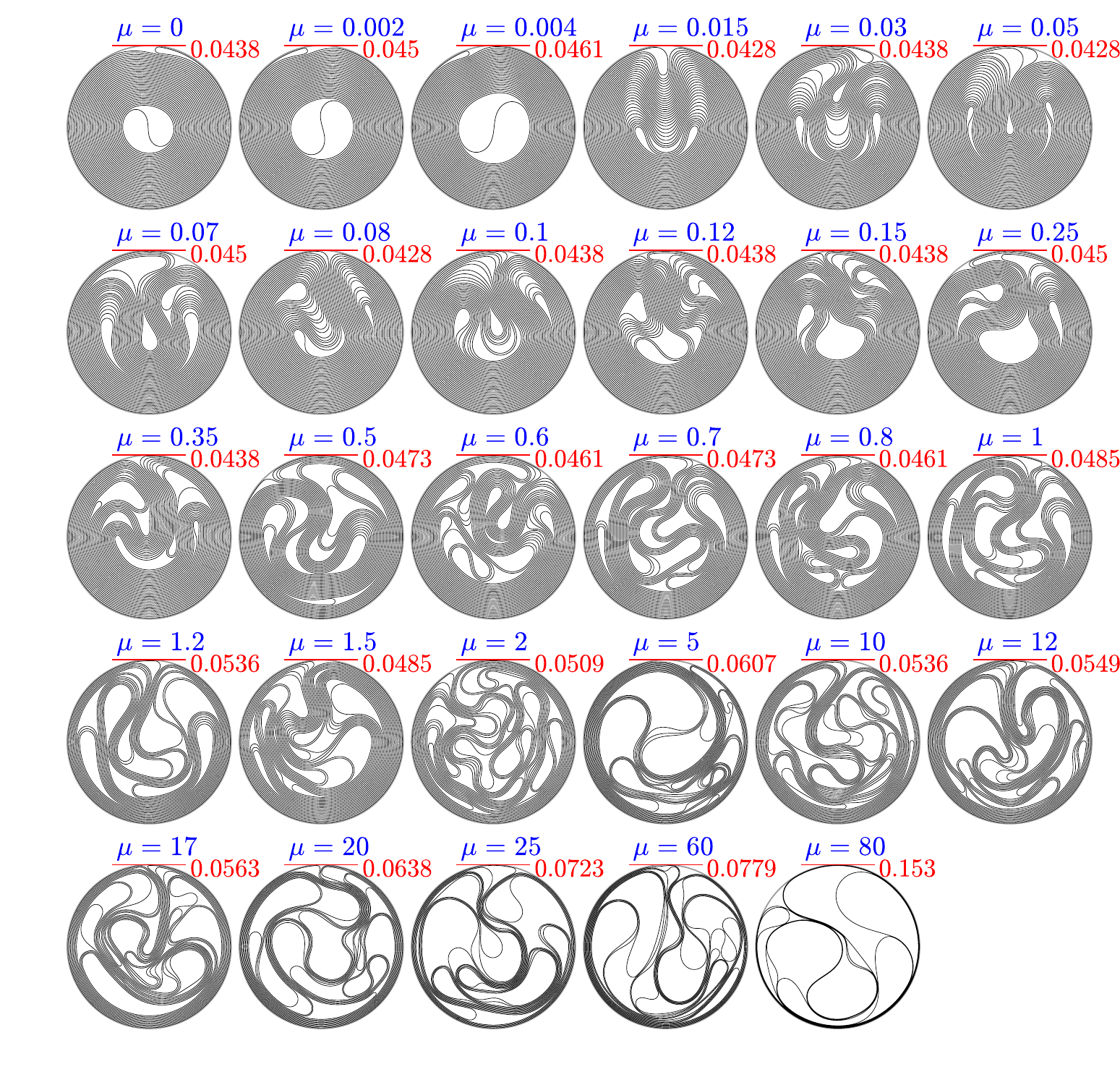}
\caption{Filament configurations with a range of friction coefficient values $\mu$ (printed in blue at the top of each configuration). Configurations are shown at the smallest $R_w$ computed (printed in red at the top and to the right of each configuration).}
\label{fig:LastShapesVsMu}
\end{figure}

For various $\mu$ in the range $[0, 80]$, we decrease $R_w$ in small steps starting from 0.82 (just before the puckered region in Fig. \ref{fig:CompareAnalyticalRing} contacts the bottom boundary) and produce a sequence of energy minimizers, as in Fig. \ref{fig:ZeroFrictionFig}. In Fig. \ref{fig:LastShapesVsMu} we show only the final configuration computed for each $\mu$ (labeled in blue above each filament), at the smallest $R_w$ of the sequence (in red at the top and right of each filament). Starting at the top left, we see spirals when the friction coefficient is very small ($0 \leq \mu \leq 0.004$). The next set of shapes ($0.015 \leq \mu \leq 0.1$) have almost as many layers as the spirals, but they have folded inward after forming many spiral turns. As $\mu$ increases above 0.1, there is increasing separation of the layers, and more sharp curves. As in the frictionless case, the computed equilibrium shape abruptly loses smoothness when the curvature reaches a threshold value $\sim 1/n$ (typically $1000-5000$), and this occurs at somewhat larger $R_w$ when $\mu$ is larger. In common materials, $0 \leq \mu \lesssim 2$, with values above 1 occurring, for example, in self-contact of certain metals and rubber tires on dry asphalt \cite{bhushan2000modern}. Sliding friction coefficients up to 21 have been observed for certain metals in a vacuum \cite{buckley1971friction,deulin2010friction}, though it is unclear if such large values can occur in static friction. Nonetheless, we consider friction coefficients as large as 80 to study the behavior in the large-friction limit, where sliding of the contacts is almost completely prevented. It appears that the trend towards decreased layering with larger $\mu$ continues up to the largest $\mu$ (comparing cases with similar $R_w$ in Fig. \ref{fig:LastShapesVsMu}, e.g. comparing $\mu$ = 1.2, 10, and 17, or comparing 5 and 20), though we do not precise quantify the trend here.

\begin{figure}[!h]
\hspace*{-0.2in}
\includegraphics[width=6.2in]{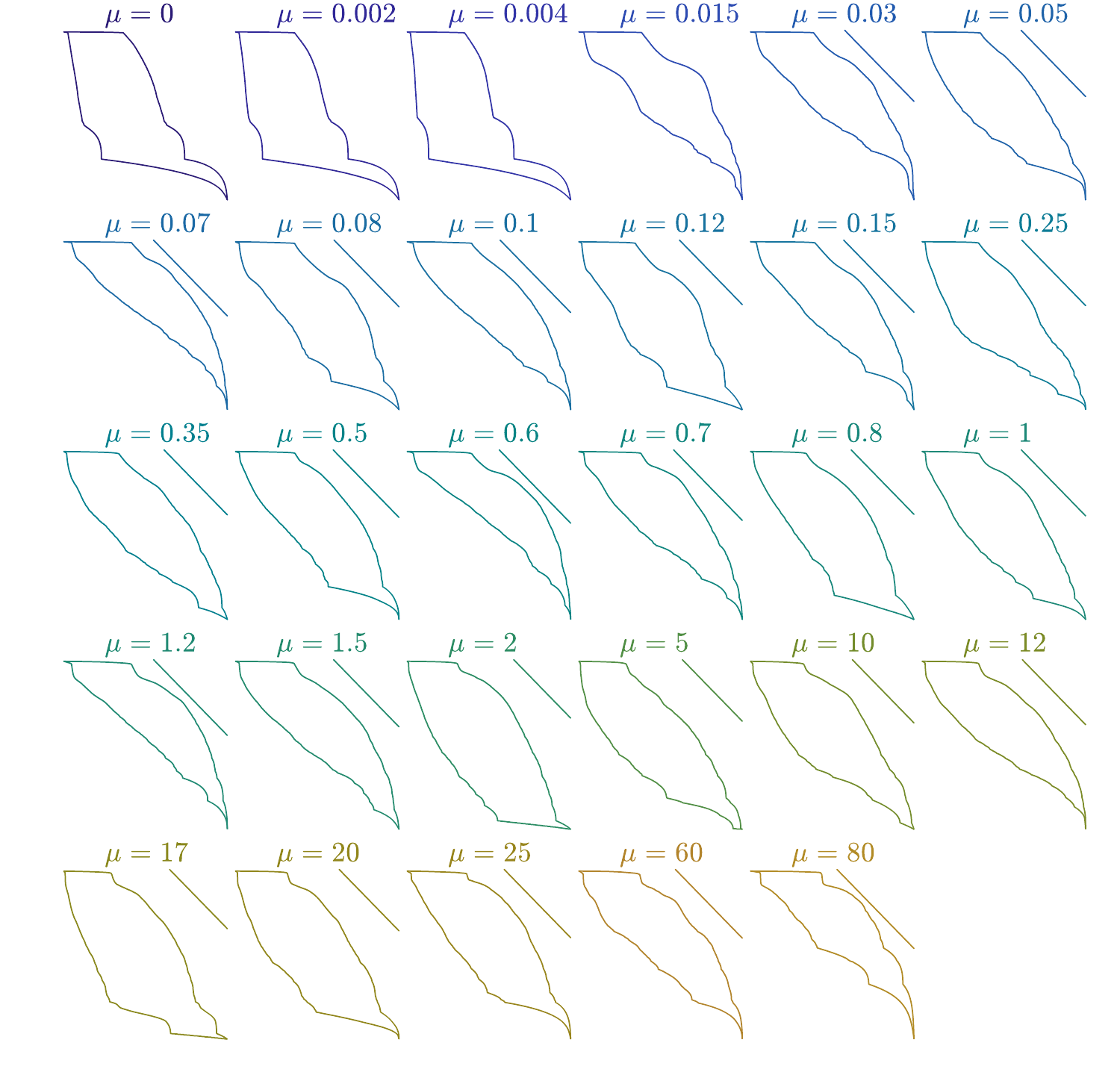}
\caption{Complementary distribution function of curvature $\bar{F}_{|\kappa|}(x)$ (defined in (\ref{CCDF})) for each of the shapes in Fig. \ref{fig:LastShapesVsMu}. For each $\mu$, the function is plotted twice: on a log-linear scale (lower curvilinear line) and a log-log scale (upper curvilinear line). For $\mu \geq 0.03$, a third line is shown at upper right: a straight line showing the scaling $x^{-1}$ on the log-log scale.}
\label{fig:KappaComplementaryDistribution}
\end{figure}

The filament configurations in Fig. \ref{fig:LastShapesVsMu} show distributions of curvature magnitudes. To minimize energy, most of the filament arc length lies in regions of curvature $\sim 1/R_w$, often with many overlapping layers. 
If self-penetration were allowed, the two lobes in the third and subsequent configurations of Fig. \ref{fig:ZeroFrictionFig} could interpenetrate, and the curvature everywhere could remain $\sim 1/R_w$ (though nonuniform). With self-penetration not allowed, for each filament in Fig. \ref{fig:LastShapesVsMu}
there are short regions where the curvature magnitude reaches a maximum $\gg 1/R_w$, and a distribution of intermediate curvature magnitudes. It is natural to ask if the distribution has a characteristic behavior, e.g. a power law, particularly at larger $\mu$ where there is a heterogeneous spatial distribution of curvatures.  In Fig. \ref{fig:KappaComplementaryDistribution} we plot the complementary distribution function of $|\kappa|$ for each filament in Fig. \ref{fig:LastShapesVsMu}. This is the function of $x$ whose value is the fraction of arc length where the curvature magnitude exceeds $x$:
\begin{align}
    \bar{F}_{|\kappa|}(x) &\equiv \frac{1}{2\pi}\mbox{measure}\left\{s: |\kappa(s)| > x \right\}. \label{CCDF}
\end{align}
\nn Thus $\bar{F}_{|\kappa|}$ decreases from 1 to 0 as $x$ increases from 0 to $|\kappa_{max}|$. For each $\mu$, we plot $\log_{10} \bar{F}_{|\kappa|}(x)$ twice in Fig.
\ref{fig:KappaComplementaryDistribution}, with
respect to $x$ (lower curvilinear line) and $\log_{10}(x)$
(upper curvilinear line). We truncate the range of $x$ to omit a horizontal asymptote at the left (lower) end and a vertical asymptote at the right (upper) end. Specifically, we set the lower limit of $x$ to $1/(8R_w)$ (one-eighth the outer boundary curvature), where 
$\bar{F}$ is at least 0.97 in all cases; below this range 
$\log_{10} \bar{F}$ is nearly a horizontal line. The upper limit $x$ is where $\bar{F}(x) = 10^{-4}$; slightly beyond this $x$, $\log_{10} \bar{F}$ has a vertical asymptote.
We omit the horizontal and vertical axes and labels to avoid visual clutter. 

For $0 \leq \mu \leq 0.004$, the graphs have three steep drops. The first corresponds to the spiral turns, which range in curvature from $1/R_w$ to about $4/R_w$ as one moves to the inside of the spiral. The second and third steep drops correspond respectively to curvature maxima in the S-curve at the spiral center, and the curvature peak in the sharp outer loop, the global curvature maximum. Near a local curvature maximum $\kappa_0$, $\kappa(s) \approx \kappa_0 + \kappa''(s-s_0)^2/2$,
so $ds/d\kappa \sim 1/\sqrt{\kappa_0-\kappa(s)}$, and the slope of $\bar{F}$ should be infinite there.
At $\mu = 0.015$, the spiral undergoes an infolding in Fig. \ref{fig:LastShapesVsMu}, so there is more of a continuous range of curvatures up to the global maximum, and consequently the curvature distributions fall off less sharply in Fig. \ref{fig:KappaComplementaryDistribution}.
For $\mu \geq 0.03$, we include at the upper right, adjacent to each graph on the log-log axes, a straight
line with the scaling $x^{-1}$. There is no clear power-law scaling that applies generally for the tail distributions, neither for the more structured deformations with $\mu \lesssim 0.5$, nor for the more disordered configurations with $\mu \gtrsim 0.5$.
Some of the log-linear plots (lower curves) are relatively straight in certain regions, but there is usually no clear exponential behavior. It is possible that a clear asymptotic behavior could emerge by going to much smaller $R_w$, but this is not feasible with the current algorithm.

\begin{figure}[!h]
\hspace*{-0.2in}
\includegraphics[width=4.5in]{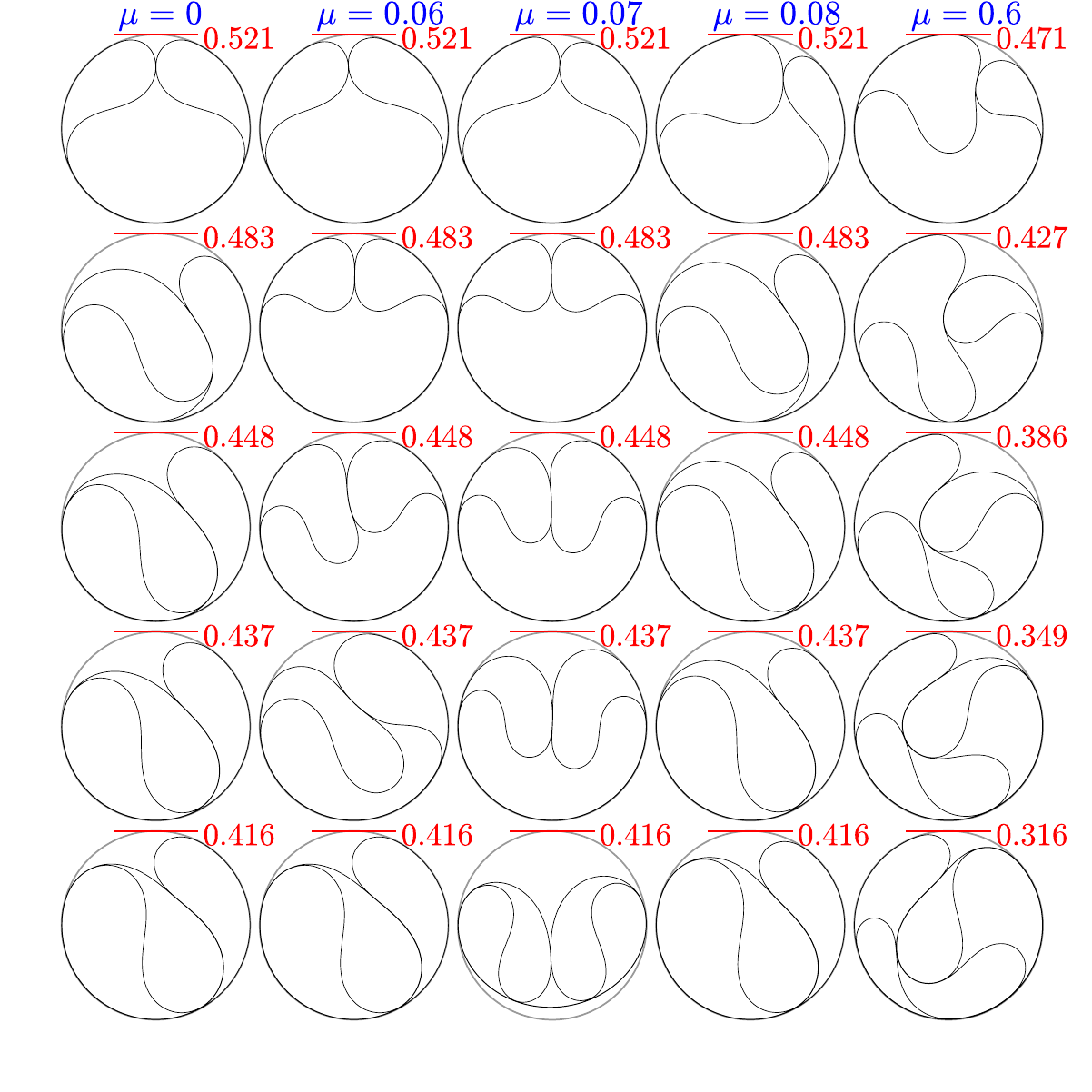}
\caption{Each column shows a sequences of filament configurations with a different $\mu$ and decreasing $R_w$, shortly after the initial symmetry-breaking.}
\label{fig:ShapeTransitions1}
\end{figure}

We now discuss some of the typical deformations that occur with friction. In Fig. \ref{fig:ShapeTransitions1},
each column shows a sequence of filament configurations for a different $\mu$, shortly after the first contact of the two symmetric lobes at $R_w$ near 0.5. For $\mu$ = 0, the left lobe slides inside the right lobe, leading to the first turn of the spiral. For $\mu$ = 0.06, friction initially prevents this relative sliding, and the lobes move inward together until, near $R_w = 0.437$, elastic forces are large enough to overcome friction and cause the left lobe to slide past the right, leading to the same configuration as the frictionless filament at $R_w$ = 0.416. For $\mu$ = 0.07, the two lobes instead remain in contact until they reach the outer boundary, leading to substantially different deformations subsequently. For $\mu = 0.08$, the two lobes are more asymmetric at $R_w = 0.521$ than in the previous cases, and this allows the left lobe to again slide past the right lobe and initiate the spiral shape. Apart from $\mu = 0.07$, the first turn of the spiral shape occurs for a range of $\mu$ up to about 0.5. For $\mu > 0.5$, a very different deformation occurs in which the left lobe is approximately bifurcated by the right lobe. An example with $\mu$ = 0.6 is shown in the last column of Fig. \ref{fig:ShapeTransitions1}.

\begin{figure}[!h]
\hspace*{-0.4in}
\includegraphics[width=6in]{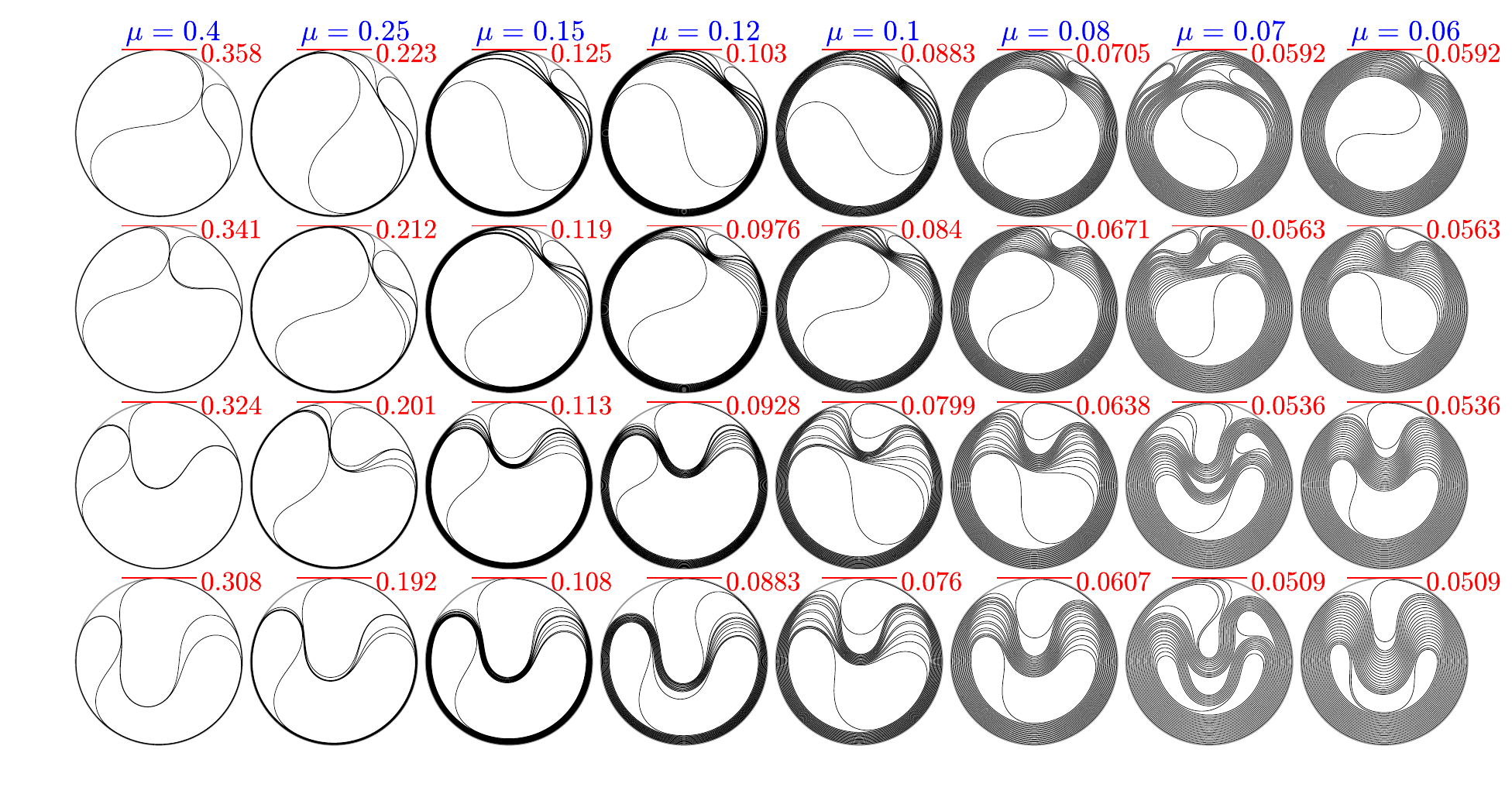}
\caption{Each column shows a sequences of shapes shortly after the first major departure from the spiraling deformation.}
\label{fig:ShapeTransitions2}
\end{figure}

For $0.015 < \mu < 0.5$ the first departure from the spiraling deformation takes on a characteristic form, shown for various $\mu$ by the sequences of snapshots in each column of Fig. \ref{fig:ShapeTransitions2}. After some number of turns of the spiral have formed (1, 3, 7, 9, 11, 14, and 18 in the first through sixth and eighth columns, respectively), the small loop outside the spiral becomes enlarged and bifurcates the spiral. The friction between the outer loop and the spiral increases with each turn, so the bifurcation occurs possibly when a critical frictional force is reached---after a number of turns that decreases with $\mu$. A bifurcation of a perturbed spiral shape occurs when $\mu = 0.07$. Despite the initial deviation from the spiral shown in Fig. \ref{fig:ShapeTransitions1}, when $\mu = 0.07$, a large portion of the filament has adopted a spiral shape at the top of Fig. \ref{fig:ShapeTransitions2}.

\begin{figure}[!h]
\hspace*{-0.2in}
\includegraphics[width=6.2in]{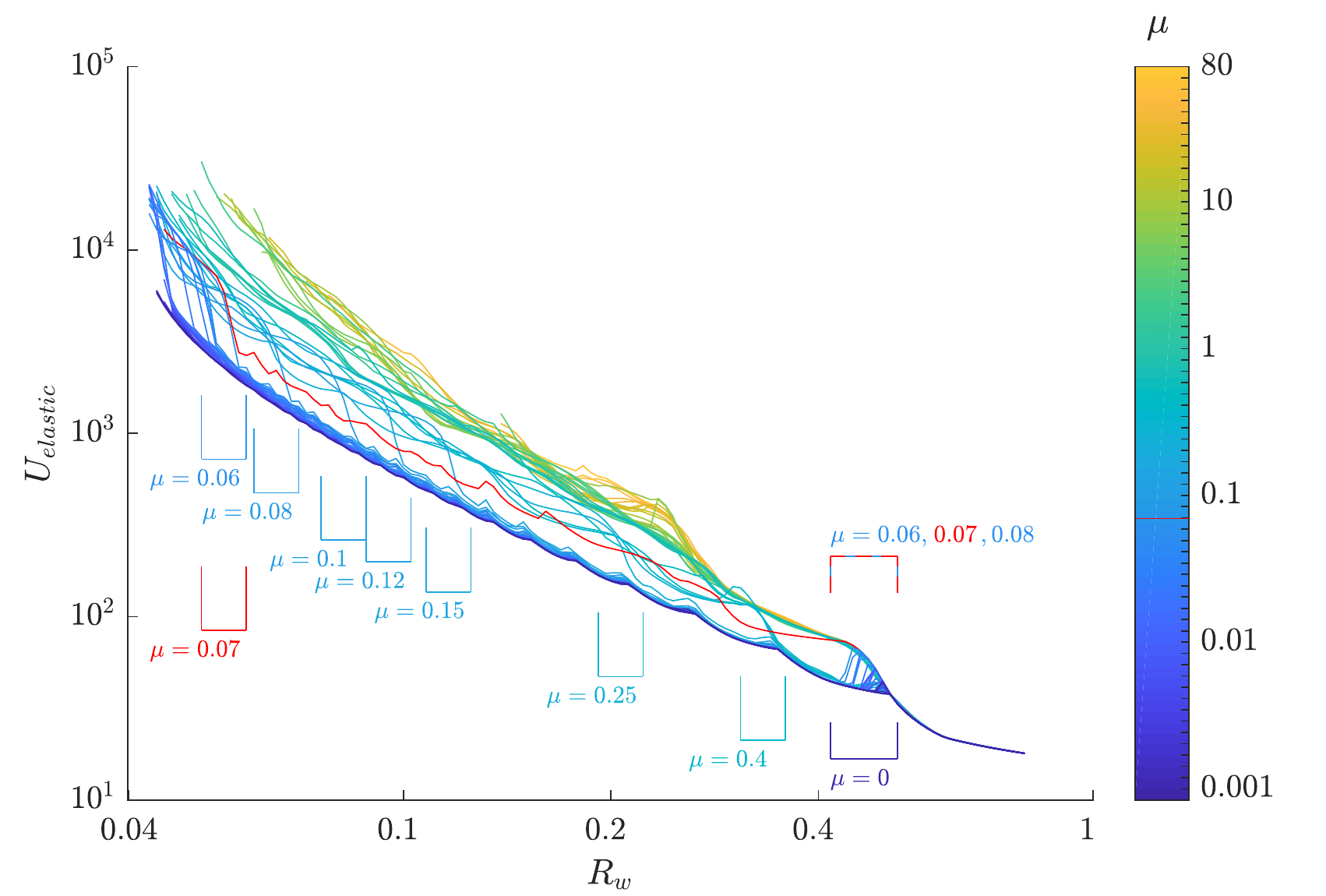}
\caption{Elastic energy versus boundary radius $R_w$. 
The values of $\mu$ are labeled by the tick marks on the color scale at right. They are 0, then the sequence \{1, 1.2, 1.5, 1.7, 2, 2.5, 3, 3.5, 4, 5, 6, 7, 8\} multiplied by $10^{-3}, 10^{-2},10^{-1},10^{0}$, and $10^{1}$, in order.
Certain ranges of $R_w$ are labeled by $\mu$ values. Those at the far right ($R_w > 0.4$) correspond to sequences of shapes shown in Fig. \ref{fig:ShapeTransitions1}. The remaining ranges correspond to the sequences in Fig.
\ref{fig:ShapeTransitions2}.}
\label{fig:UElasticVsRw}
\end{figure}

The effect of the bifurcations can be seen clearly in the evolution of the elastic energy, shown in Fig. \ref{fig:UElasticVsRw} for 66 values of $\mu$ indicated by the tick marks on the color scale at right. Although the tick marks are equispaced, this is only approximately a logarithmic scale; the precise values of $\mu$ at the tick marks are given in the figure caption. The lowest curve in the figure has $\mu = 0$ (darkest blue line), and follows a slightly scalloped shape as it moves from right to left, with decreasing $R_w$. The same curve was shown in Fig. \ref{fig:ZeroFrictionAnalysis}A (solid blue) to follow
an $R_w^{-2}$ scaling until thickness effects become important. The bracket below the curve labeled $\mu = 0$ at $R_w$ slightly greater than 0.4 shows the range of values in Fig. \ref{fig:ShapeTransitions1}. A corner occurs in the energy plot as one lobe passes inside the other. The red and blue dashed bracket above the curve labels the same range of $R_w$ for the other $\mu$ in Fig. \ref{fig:ShapeTransitions1}, with the anomalous case $\mu = 0.07$ in red in Fig. \ref{fig:UElasticVsRw}. The energy curves for $\mu$ = 0.06 and 0.08 at first rise sharply like the red curve as $R_w$ decreases through the bracketed range, but then drop sharply, back to the $\mu = 0$ curve, as one lobe moves inside the other. The red curve stays high, as do other curves at much larger $\mu$ ($\geq$ 0.5), but it eventually drops to an intermediate position among the curves as the $\mu = 0.07$ case assumes a perturbed spiral shape. Most of the other curves with $0 \leq \mu < 0.5$ have a sharp drop (like 0.06 and 0.08) as one lobe passes inside the other. Following these curves (light blue) to smaller $R_w$, they also have a scalloped shape as more spiral turns are formed, until they abruptly jump to larger energies, when the spiral bifurcates after a certain number of turns. These jumps are indicated by brackets below the energy curves in Fig. \ref{fig:UElasticVsRw}, and correspond to the sequences of shapes in Fig. \ref{fig:ShapeTransitions2}. The general pattern is that the energy (and deformations) approximate the $\mu = 0$ case until the two lobes meet, for 
$\mu \geq$ 0.5, or until the spiral bifurcates, for
$0 < \mu < 0.5$. In all cases, the energies subsequently follow an $R_w^{-2}$ scaling approximately, though with prefactors up to about 8 times greater than in the $\mu = 0$ case. For each filament in Fig. \ref{fig:LastShapesVsMu}, the majority of the arc length has curvature comparable to $1/R_w$, but some has much larger curvature and energy, and this portion becomes larger as $\mu$ increases.

\begin{figure}[!h]
\hspace*{-0.2in}
\includegraphics[width=6.2in]{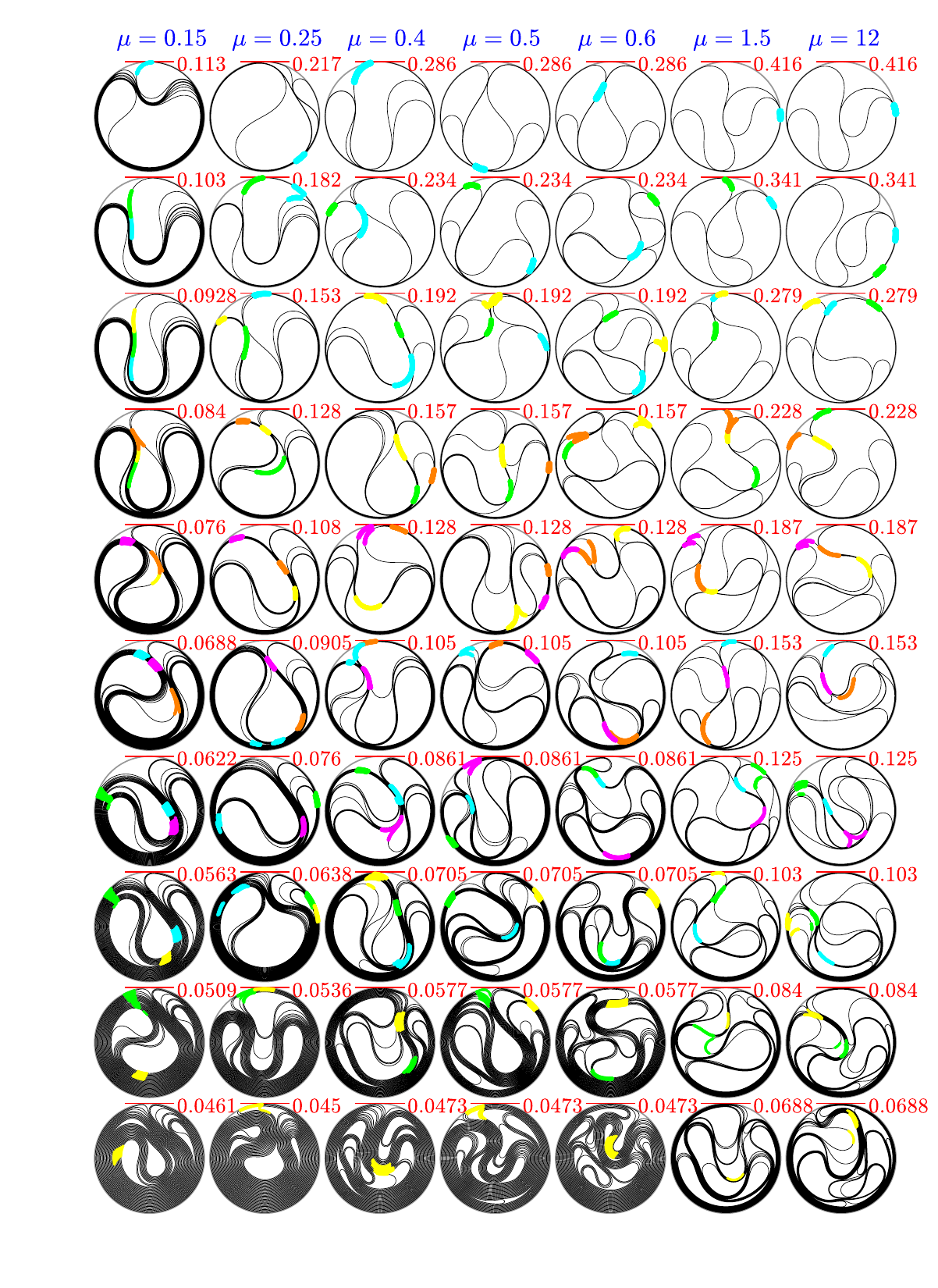}
\caption{Each column shows a sequences of shapes leading to the final computed shapes, with combinations of spiraling deformations (more common at smaller $\mu$, left), and bifurcations (more common at larger $\mu$, right).}
\label{fig:ShapeTransitionsMu01To06}
\end{figure}

We now investigate filament deformations along the upper branches of the energy curves, after the initial deviation from the spiral configuration. Fig. \ref{fig:ShapeTransitionsMu01To06} shows seven cases. The first three columns have lower values of $\mu$, and initially form spirals. For $\mu = 0.15$, the first few snapshots show the bifurcation of the initial spiral. Certain ranges of points on the filament are given the same color (e.g. blue, green, yellow) over three successive snapshots, to give a sense of the deformation sequence. After the bifurcation completes ($R_w$ = 0.084), a second spiraling deformation occurs, and continues until the smallest $R_w$, 0.0461. A similar phenomenon occurs in the second column: the initial bifurcation completes at 
$R_w$ = 0.153, a second spiraling deformation occurs until
$R_w$ = 0.0638, and then a second bifurcation occurs at
$R_w$ = 0.0536 and 0.045. In the third column, the first bifurcation ends near the first frame, $R_w$ = 0.286, a second spiraling deformation occurs and ends with a second bifurcation at $R_w$ = 0.128 and 0.105, and then a third spiraling deformation occurs and ends with a third bifurcation in the last frame, $R_w$ = 0.0473. The fourth and fifth columns begin similarly to each other, with one of the initial lobes bifurcating the other instead of a spiral deformation, as shown in the last column of Fig. \ref{fig:ShapeTransitions1}. Two new lobes meet at the tops of the first frames of the fourth and fifth columns of Fig. \ref{fig:ShapeTransitionsMu01To06}. In the fourth column, the larger lobe (right) has slid inside the smaller lobe at
$R_w$ = 0.234. This lobe is then bifurcated by the smaller lobe, until 0.128. A spiraling motion then occurs on the right side of the filament (following the pink dots from 0.128 to 0.0861), followed by a bifurcation (following the blue dots from 0.0861 to 0.0705) and another bifurcation between the last two frames. The fifth column proceeds differently than the fourth: the top left lobe (blue dots) bifurcates the right, and is then bifurcated by a curve on the right (green dots). The flat region adjacent to the green dots is bifurcated by the curves with orange dots (0.157 to 0.105). Next, the curve with blue dots bifurcates the rest of the filament (0.105 to 0.0705). In the last three frames, different parts of the filament fold into S-shapes (e.g. near the yellow dots). In the sixth and seventh columns the dynamics are broadly similar to the fifth: sharp curves repeatedly form and invade flatter regions on the rest of the filament, resulting in a series of bifurcations. Because the filament is under compression, curved regions are similar to arches in bridges and buildings, and can resist forces of magnitude $|T||\kappa|$, where $|T|$ and $|\kappa|$ are the magnitudes of the compression and curvature, respectively. When friction is large, curved regions cannot slide along flatter surfaces, and instead tend to bifurcate them, eventually being resisted by contact with the outer boundary.

\begin{figure}[!h]
\hspace*{-0.2in}
\includegraphics[width=6.2in]{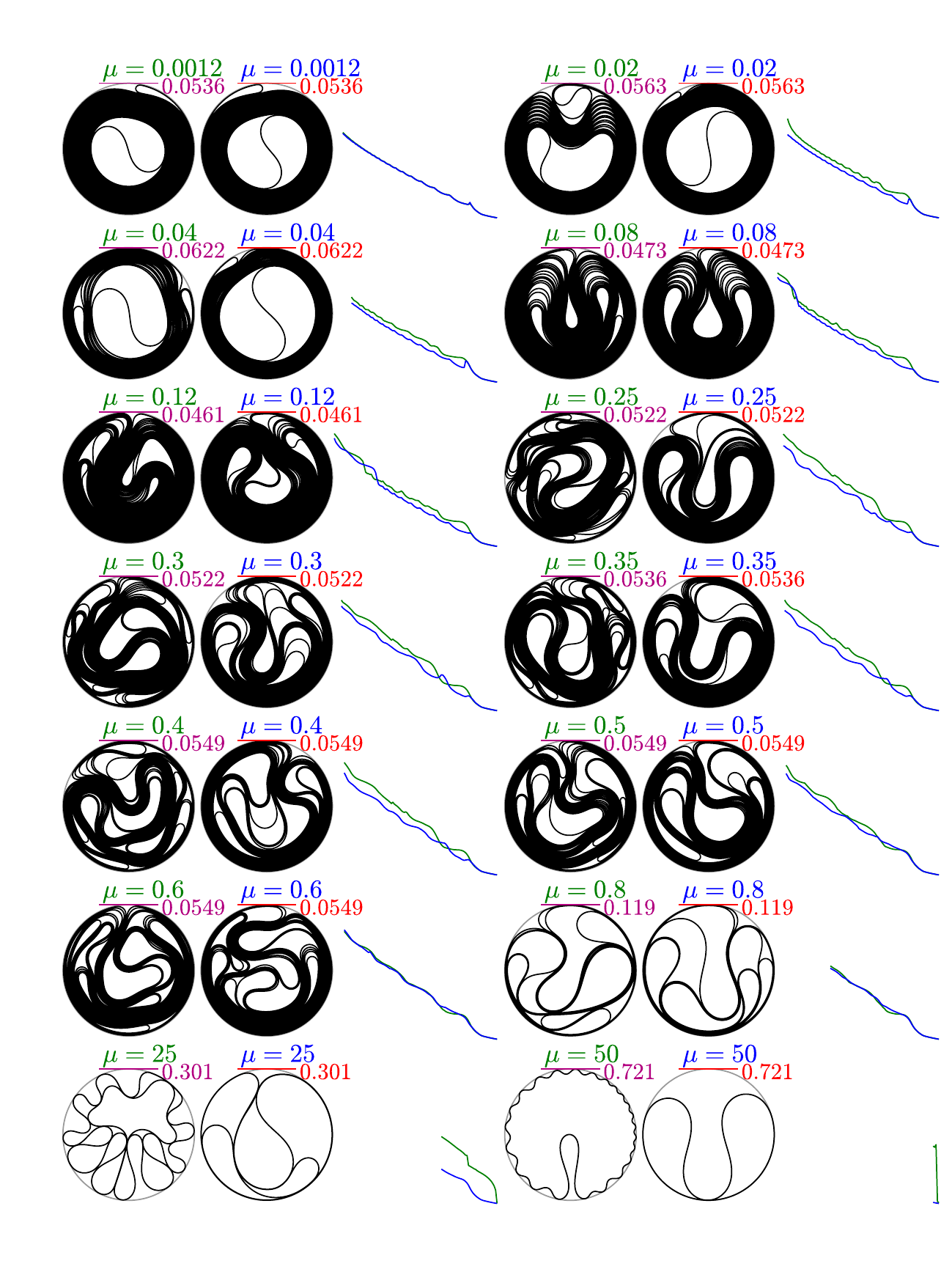}
\caption{Comparison of configurations with wall friction (left member of each pair) and without (right member). To the right of each pair are the corresponding graphs of elastic energy (green with wall friction, blue without).}
\label{fig:LastShapesVsMuCompareWall}
\end{figure}

So far, we have assumed that the filament has friction at self-contacts, but not with the wall. We now briefly consider wall-filament friction (with the same coefficient as filament-filament friction), by updating $U_{wall}$ in (\ref{Uwall}) to (\ref{Uself1}), with the appropriate $s_i$ and $n_i$. We find that in general the energy minimization algorithm is somewhat less robust with wall friction. In
most cases with $\mu \lesssim 0.6$, it computes minimizers successfully to about the same minimum $R_w$ as without wall friction, but in some cases it fails to find an energy minimizer much earlier. Early breakdown is more common for $\mu  \gtrsim 0.6$. One possible explanation is that when $R_w$ is decreased, the wall is brought closer to the filament, creating a mismatch in radial force on the filament there. So the filament tends to move inward. If a filament segment lying entirely along the wall moves inward, its arc length decreases, which greatly increases stretching energy. Without wall friction, the filament arc length is restored by sliding tangentially along the wall. With wall friction, this sliding is resisted, and it is potentially more difficult to find an energy minimizing state. 

In Fig. \ref{fig:LastShapesVsMuCompareWall} we compare pairs of configurations with and without wall friction, on the left and right side of each pair, respectively. The graphs of elastic energy for the two cases are shown to the right of the respective pair (in green with wall friction, and in blue without; the axes are omitted, but are the same as in Fig. \ref{fig:UElasticVsRw}). At $\mu = 0.0012$, friction is small enough that the two results are spirals, almost the same. At $\mu = 0.02$, the wall friction case follows a higher-energy branch early on, via the same configuration as $\mu = 0.07$ at the bottom of Fig. \ref{fig:ShapeTransitions1}, and eventually forming a perturbed spiral shape with an S-curve adjacent to the wall instead of a single sharp bend as for the spiral. At $\mu = 0.04$ in the wall friction case, the same initial stages occur as for 0.02, but the outer S-curve splits into two separate sharp bends and move apart during the deformation. At $\mu = 0.08$ and 0.12, the cases with and without wall friction are similar at late stages, both predominantly spiral deformations, with a single bifurcation at small $R_w$. The wall friction cases generally have higher energy at a given $R_w$, but not always. For $0.3 \leq \mu \leq 0.5$, the two cases are again qualitatively similar, with elastic energy somewhat larger and sharp bends somewhat more numerous with wall friction. At $\mu = 0.6$ and 0.8 the two energy curves are almost identical, and both filaments' energies follow the higher branch. In the last row, $\mu$ is much larger, and there is a qualitative difference between the two cases. Wall friction is so large that tangential sliding at the wall is almost completely prevented, and instead the filament buckles inward in series of small undulations. These deformations lead to a very different evolution in the earliest stages of the deformation.





\section{Conclusion}
We have used an X-lattice spring model to study the packing of thin elastic rings with friction inside a contracting circular boundary. The model agrees well with solutions of the inextensible elastica equation for a uniformly loaded cantilever and in the initial stages of puckering by the elastic rings. In the frictionless case, the X-lattice model adopts a spiral configuration with a single sharp curve trapped at the outer wall. A simple model predicts that the curvature of the sharp curve scales with wall radius $R_w$ as $R_w^{-3/2}$, and the total elastic energy scales as $R_w^{-2}$, with next-order terms smaller by powers of $R_w^{1/2}$. The numerical solutions agree well with the first two terms in the expansion. 

With nonzero friction, the distribution of curvature becomes more heterogeneous during the packing. By plotting the complementary curvature distribution function we found that large curvatures are much less common than small curvatures, as expected given the elastic energy density $\kappa^2$. The curvature distribution functions do not clearly fit a power law or exponential scaling with $R_w$ for the range $0.04 < R_w < 1$ computed here, but such scalings may emerge when $R_w$ is decreased further. 

In many cases the deformation process can be decomposed into two main types of deformations: spiraling and bifurcations, which occur in an alternating fashion. In spiraling deformations, part of the filament slides tangentially and rotates with respect to the remainder. When friction is large enough, spiraling is prevented and instead a sharp curve on the filament moves into the remainder mainly in the direction normal to the contact, and bifurcates it. If the friction coefficient is not too large ($\mu \lesssim 0.5$), spiraling deformations may occur repeatedly. For $\mu \gtrsim 0.5$, the deformation is mainly a sequence of bifurcations with little relative tangential sliding at contacts. The deformations are qualitatively similar with and without wall friction up to $\mu \approx 1$. At larger $\mu$, the wall friction case is dominated by buckling near the wall.

\begin{acknowledgments}
This research was supported by the NSF Mathematical Biology program under
award number DMS-1811889.
\end{acknowledgments}

\appendix
\section{Relating the spring stiffness $k_1$ to the bending modulus $B$ \label{k1}}

We relate $k_1$ to $B$ by considering deformations of the X-lattice in the continuum limit $h \to 0$. In this limit, the points on the lattice converge to a curve in the plane. If the curve has zero curvature (is straight), the undeformed lattice can converge to it. More generally, the curve has nonzero curvature, and then the lattice must be deformed in the continuum limit. With a particular type of deformation---that assumed by the Euler-Bernoulli model of beam bending---the lattice can converge to a given curve with a strain that is $O(h)$ in the top and bottom springs and $O(h^2)$ in the remaining springs. Let us assume that the centerline of the lattice---the points that are the average of the top and bottom rows of points (connected by the nearly vertical springs in Fig. \ref{fig:CantileverFig}A for example)---converges to a curve with local curvature $\kappa$. The strain in the top and bottom springs is then $\pm h\kappa/2$, while the strain in the side springs (pink in Fig. \ref{fig:CantileverFig}A) and the diagonal springs is $O(h^2)$ (see Fig. \ref{fig:CantileverFig}D). The total elastic energy per unit length of the top and bottom springs with strain $\pm h\kappa/2$ is
\begin{align}
    2\frac{k_1}{2}\left(\frac{h\kappa}{2}\right)^2 h.
\end{align}
The elastic energy per unit length of an Euler-Bernoulli beam is $B\kappa^2/2$. Matching these expressions we have $B = k_1 h^3/2$. The same result can be obtained using the formula $B = EI$, with $E$ the Young's modulus and $I$ the area moment of inertia of a beam cross-section. We compute $I$ assuming that all the material in the cross-section is concentrated in layers of thickness $\eta$ at the top and bottom springs, located $\pm h/2$ from the centerline. Then we have
\begin{align}
    B = EI = EW\eta h^2/2 = k_1 h^3/2.
\end{align}
\nn The last equality follows by writing the elastic energy per unit length in the top or bottom spring assuming it has strain $\epsilon$. In the continuum model it is $EW\eta \epsilon^2/2$, while in the discrete model it is $k_1 h \epsilon^2/2$, using (\ref{Uelastic}) with $d_{ij} = h$.

\section{Solving for the shapes of puckered rings \label{pucker}}

We integrate a version of the elastica equation obtained from the normal and tangential components of (\ref{cantilever}) without external loading:
\begin{align}
    -\partial_{ss}\kappa + T\kappa = 0 \quad ; \quad T(s)  = T_1 + \frac{1}{2}(\kappa_1^2 - \kappa^2). \label{elastica}
\end{align}
\nn Here $T_1$ and $\kappa_1$ are the values of tension and curvature at $s_1$, where one end of the puckered region meets the outer boundary. It is convenient to rescale the problem geometry so that the outer boundary has fixed radius 1 and the puckered ring has a range of lengths $\geq 2\pi$. Each solution in Fig. \ref{fig:CompareAnalyticalRing} corresponds to a different choice of $\partial_s\kappa$ at $s_1$. To start the Runge-Kutta integration of (\ref{elastica}), we set $\kappa_1 = 1$ and use a range of values of
$\partial_s\kappa_1 > 0$. For each value of $\partial_s\kappa_1$, we use a range of guesses for $T_1$. By the symmetry of the puckered region, we know that its far end is reached at the first location where $\kappa = 1$ and $\partial_s\kappa < 0$. We then integrate $\kappa$ starting from a point on the unit circle ($\mathbf{X}_1 = [1, 0]^T$, $\theta_1 = \pi/2$) to obtain the position of the puckered region. Then the correct value of $T_1$ may be identified: that for which the far end of the puckered region also lies on the unit circle. We thus obtain the full family of puckered regions (including those in Fig. \ref{fig:CompareAnalyticalRing}).


\bibliographystyle{unsrt}
\bibliography{ElasticRing}

\end{document}